\documentclass{emulateapj}
\usepackage[english]{babel}
\usepackage{graphicx}
\usepackage{hyperref}
\usepackage{subfigure}
\usepackage[toc,page]{appendix}
\usepackage{color}
\usepackage{ulem}

\begin{document}

\title{Evolution of Supernova Remnants near the Galactic Center}

\author{A. Yalinewich\altaffilmark{1}, T. Piran\altaffilmark{1} and R. Sari\altaffilmark{1}}
\altaffiltext{1}{Racah Institute of Physics, the Hebrew University, 91904, Jerusalem, Israel}

\date{\today}

\begin{abstract}
Supernovae near the galactic center evolve differently from regular galactic supernovae. This is mainly due to the environment into which the supernova remnants propagate. Instead of a static, uniform density medium, SNRs near the galactic center propagate into a wind swept environment with a velocity away from the galactic center, and a graded density profile. This causes these SNRs to be non - spherical, and to evolve faster than their galactic counterparts.
We develop an analytic theory for the evolution of explosions within a stellar wind, and verify it using a hydrodynamic code. We show that such explosions can evolve in one of three possible morphologies. Using these results we discuss the association between the two SNRs (SGR East and SGR A's bipolar radio/X-ray Lobes) and the two neutron stars (the cannonball and SGR J1745-2900) near the galactic center. We show that, given the morphologies of the SNR and positions of the neutron stars, the only possible association is between SGR A's bipolar radio/X-ray Lobes and SGR J1745-2900. The compact object created in the explosion of SGR East remains undetected, and the SNR of the supernova that created the cannonball has already disappeared.
\end{abstract}

\section{Introduction}

The galactic center (GC henceforth) is an active star forming site. The star formation rate in the GC is $3 \cdot 10^{-3} \, M_{\odot} / \rm{year}$ \citep{figer_et_al_2004}. The abundance of young and short lived massive stars is higher near the GC, compared with the rest of the Galaxy. These stars influence their environment in two ways. First, they emit stellar winds \citep{1997A&A...325..700N}. The  mass loss due to winds is $3 \cdot 10^{-3} M_{\odot}/\rm{year}$, and these winds propagate at a velocity of about $700 \, \rm{km/s}$ \citep{rockefeller_et_al_2004}. Second, a supernova occurs close to the GC every $10^4$ years \citep{zubovas_et_al_2013}. When a supernova occurs, the shock wave does not propagate into a stationary, constant density medium, as in the rest of the galaxy, but into a moving medium with a graded density profile.

This motivates us to develop an analytic theory for an explosion within a wind. We verify our analytic results using a state of the art hydrodynamic simulation RICH \citep{Yalinewich_2015}. Our simplified model neglects a lot of details in the GC, but we show that these do not affect our results.

Two SNRs have been detected near the GC: SGR East \citep{mezger1989continuum,baganoff_et_al_2003} and SGR A's bipolar radio/X-ray Lobes \citep{ponti_et_al_2015}. Two neutron stars have also been detected close to the GC: CXOGC J174545.5-285829 (a.k.a. the Cannonball) \citep{nynka_et_al_2013} and SGR J1745-2900 \citep{kaspi_et_al_2014}. The cannonball is thought to be a relic from the same supernova that created SGR East, and SGR J1745-2900 is thought to be the relic from the supernova that created the SGR A's bipolar radio/X-ray Lobes. 

The associations between SNRs and NSs is largely based on the assumption that SNRs near the GC evolve as regular SNRs. 
However, \citet{rockefeller_et_al_2005} studied the evolution of SGR East and showed that it is much younger than previously thought. \citet{rimoldi_et_al_2015} studied the evolution of general SNRs near a general GC. They showed that the explosion can be highly non spherical. The reason for that is that fluid elements moving toward the center decelerate faster than fluid elements moving away. They assumed a graded density profile, but with zero velocity (stationary medium). We generalize the results of these two works, and reexamine the implication on the association between the SNRs and NSs.

This paper is organized as follows. In section \ref{sec:analytic_estimates} we consider an idealized analytic model for an explosion within a wind. In section \ref{simulations} we verify our analytic results using a numerical simulations. We show that despite having neglected many objects and phenomena that arise in the GC, our simplified model still captures the important features of the problem. In section \ref{sec:snr_gc} we reexamine the association between the SNRs and SNs in light of ours results from the previous sections. Finally, in section \ref{sec:discussion} we summarize the results.

\section{Analytic Estimates} \label{sec:analytic_estimates}

We consider a simplified analytic model for the problem. We begin with a very basic picture of the GC, and gradually introduce details. In the first stage, we consider just the wind. We discuss the density profile from a point - like wind source, and also the density profile inside a spherical region where the wind is uniformly generated (henceforth, the wind generating zone). In the second stage we consider an explosion that occurs in the same place as the point like wind source. In the third stage we allow the explosion to occur at a finite distance from the wind source. In the fourth stage, we consider an explosion near a wind generating zone of a finite radius.

\subsection{A Wind}
A spherically symmetric, cold wind emanating from a point source with mass loss rate $\dot{M}$ at a constant velocity $v$ will have the following density profile
\begin{equation}
\rho = \frac{\dot{M}}{4 \pi v r^2}. \label{eq:cold_wind_density_outer}
\end{equation}

If the wind is not emanating from a point source, but rather from a sphere of radius $r_w$, then outside $r>r_w$ the density distribution would be the same as equation \ref{eq:cold_wind_density_outer}. If the wind is generated at a uniform emissivity $\frac{3 \dot{M}}{4 \pi r_w^3}$, then inside the wind generating zone $r<r_w$ the density is given by

\begin{equation}
\rho = \frac{\dot{M} r}{4 \pi r_w^3 v}.
\end{equation}

\subsection{A Concentric Explosion} \label{sec:concentric_explosion}
We consider, first, an explosion that occurs exactly at the point like source of the wind. This problem of an explosion inside a wind blown bubble has been considered in the past in the more common scenario of a supernova inside a wind bubble blown by its own progenitor star \citep{dwarkadas_2005, medvedev_loeb_2013}. 

Here we recount the main results. We denote the energy of the explosion by $E$ and the mass of the ejecta by $M_e$. The initial speed of the ejecta is therefore $v_e \approx \sqrt{E/M_e}$, and we assume that it is much larger than the wind velocity $v_e \gg v$. The adiabatic evolution of this explosion has three stages. The first is the ejecta dominated phase. The ejecta moves at a constant velocity, and the wind is practically stationary. This stage ends when the swept up mass equals the ejecta mass. This occurs when the radius of the explosion is

\begin{equation}
r_{st} \approx \frac{M_e}{\dot{M}} v \simeq 5  \, M_{e, \odot} \dot{M}_{-3}^{-1} v_3 \, {\rm{pc}}  \label{eq:st_radius}
\end{equation}
and its age is
\begin{equation}
t_{st} \approx \frac{v M_e^{3/2}}{\dot{M} \sqrt{E}} \simeq 10^{3}  \, M_{e,\odot}^{3/2} \dot{M}_{-3}^{-1} v_3 E_{51}^{-1/2} \, {\rm{year}}
\end{equation}
where $M_{e,\odot} = M_e/M_{\odot}$, $\dot{M}_{-3} = \dot{M}/10^{-3} \, \frac{M_{\odot}}{\rm{year}}$, $v_3 = v/10^3 \rm \frac{km}{s}$ and $E_{51} = E/10^{51} \, \rm erg$.

After that, the explosion evolves as a Sedov - Taylor explosion in a graded density profile. The relation between the shock radius and the age of the supernova is given by

\begin{equation}
R \approx \left( \frac{E v}{\dot{M}} \right)^{1/3} t^{2/3}.
\end{equation}
This is somewhat different from a Sedov Taylor explosion in a uniform density medium, where $R \propto t^{2/5}$. This stage ends when the supernova energy becomes equal to the energy of the wind ejected since the explosion. At this moment the shock front velocity is comparable to the wind velocity. This occurs when the radius is

\begin{equation}
r_{dr} \approx \frac{E}{\dot{M} v} \simeq 50 E_51 \dot{M}_{-3} v_3^{-1} \, {\rm pc}
\end{equation}
and the explosion age is
\begin{equation}
t_{dr} \approx \frac{E}{\dot{M} v^2} \simeq 5\cdot 10^{4} \, E_{51} \dot{M}_{-3}^{-1} v_3^{-2} \, {\rm year}
\end{equation}

In the final stage, the evolution is still dictated by conservation of energy and mass, but the shock front is only slightly faster than the wind. Expanding $1 - \frac{v t}{R} \rightarrow 0$ yields
\begin{equation}
R \approx v t + \frac{E}{\dot{M} v} \ln \left( \frac{t}{t_0} \right )
\end{equation}
where $t_0$ is a constant.\vspace{5mm}

\subsection{An Off Centered Explosion}

Consider, now, an explosion that occurs at an initial distance $r_s$ from a point - like wind source. As the explosion expands, its velocity decreases. The qualitative behavior of the shock depends on whether the velocity of the shock drops below the value of the wind velocity before it travels a distance comparable to the distance to  the wind source. Therefore, for $r_s<r_{dr}$, the shock velocity remains larger than the wind velocity and the shock engulfs the wind source, while, for $r_s>r_{dr}$ it will be swept away by the wind.

We are interested in determining the minimal distance between the shock wave and the wind source. We consider two extreme cases. In one case, the explosion is so weak that it barely expands before being swept away by the wind. In this case the minimal distance is very close to $r_s$. In the other extreme, we can think of an explosion that is so powerful that the shock wave is only halted at distances much smaller then $r_s$. In this limit, the minimal distance between the explosion and the wind source is given by the condition that the ram pressure of the wind balance the thermal pressure of the explosion
\begin{equation}
r_b \approx r_s \sqrt{\frac{\dot{M} v r_s}{E}} \, .
\end{equation}
For an arbitrary value of the explosion energy $r_b$ will be between those two limits.

\subsection{The Wind Generating Zone}

Instead of a point - like wind source, we turn now to a case where the wind is uniformly generated inside a sphere of radius $r_w$. 
This difference will only affect the shock wave if $r_b<r_w$, otherwise, the explosion has no way of ``knowing'' that the wind emanates from a sphere rather than from a point source. In the limit of a very strong explosion, the critical separation below which the explosion penetrates the wind generating zone is
\begin{equation}
r_p \approx \sqrt[3]{\frac{E r_w^2}{\dot{M} v}}
\end{equation}

If a shock reaches significantly past the edge of the wind generating zone, then it will also cross it. Hence, such an explosion is qualitatively different from the two explosion types mentioned above. 

\begin{figure}
\includegraphics[width=0.9\columnwidth]{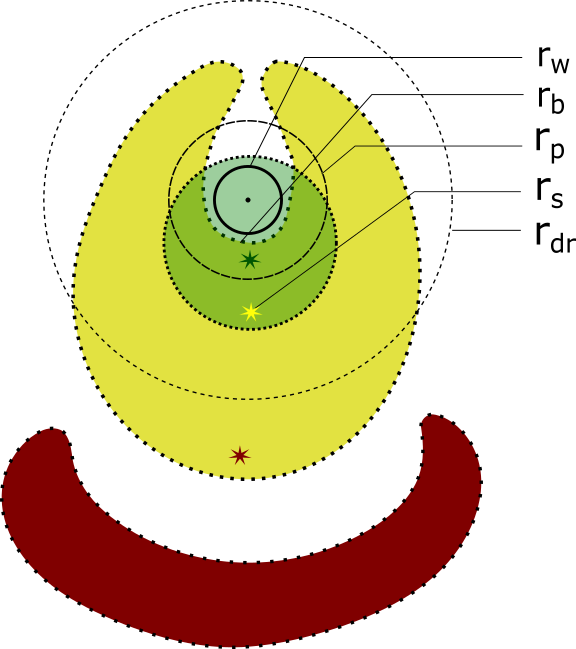}
\caption{
A schematic illustration of the three possible explosion morphologies: (i) An overwhelmed explosion where the shock wave never reaches the wind source (red); (ii) An engulfing explosion (yellow) that engulf but does not penetrate the wind generating zone; (iii) A penetrating explosion (green). The location of the progenitor is marked by a star of the same color.
The black dot marks the SMBH at the center of the wind generating zone.  $r_w$ (a solid line) denotes the radius of the wind generating zone.  $r_s$ and $r_b$ denote the position of the explosion and the minimum distance between the shock and the center for the engulfing explosion (yellow). $r_p$ (a long-dashed circle) denotes the radius of the region inside which an explosion will penetrate the wind generating zone.  $r_{dr}$  (a short-dashed circle) marks  the radius of the region inside which an explosion will engulf the wind generating zone.
\label{fig:schematic}
}
\end{figure}

In conclusion, an explosion near a wind source can have one of three possible morphologies, depicted in figure \ref{fig:schematic}. If $r_s>r_{dr}$, the explosion is overwhelmed by the wind (red crescent shape in the figure), and is swept away. If $r_p<r_s<r_{dr}$ then the explosion will engulf the wind source (yellow horseshoe shape in the figure). If also $r_s<r_p$, then the explosion will not just engulf the wind generating zone, but also penetrate it (green circle in the figure).

\subsection{Dimensionless Variables}

The problem at hand can be described in terms of three dimensionless variables. For that, we first recount the dimensional parameters of the problem (see table \ref{table:dimensional_constant}). The wind is characterized by three: the mass loss rate $\dot{M}$, velocity $v$ and the radius from which it is emitted $r_w$. The supernova is characterized by two: the energy $E$ and the ejecta mass $M_e$. Both can be combined to form the ejecta velocity $v_e \approx \sqrt{E/M}$. A sixth and final parameter is $r_s$, the separation between the explosion and the wind source. According to the $\Pi$ theorem, we can reduce the parameters describing the system to just three dimensionless constants, although the choice is not unique. The dimensionless parameters we chose are

\begin{equation}
\varpi = \frac{v_e}{v},
\end{equation}

\begin{equation}
\xi = \frac{\dot{M} r_s v}{E}
\end{equation}
and

\begin{equation}
\psi = \frac{r_w}{r_s} \, .
\end{equation}

We note that for the astrophysical problem at hand the most important from those three is $\xi$. In the GC the ejecta velocity is much larger than the wind velocity, and hence we will consider here only the case in which  $\varpi \gg 1$. Furthermore, as long as the shock wave does not penetrate the wind generating zone, $\psi$ does not have any effect.

We can now describe the main results in terms of the dimensionless variables.
The ratio $r_b/r_s$ is a function only of $\xi$. In one extreme case
\begin{equation}
\lim_{\xi\to 0} \frac{r_b}{r_s} = 1
\end{equation}
while in the other extreme
\begin{equation}
\lim_{\xi\to\infty} \frac{d \ln \left( r_b/r_s\right)}{d \ln \xi} = \frac{1}{2} \, .
\label{eq:rb_vs_xi}
\end{equation}
We will later use a computer simulation to determine the behavior of $r_b/r_s$ for intermediate values of $\xi$.

If the explosion occurred outside the wind generating zone, it engulfs the wind source if $\xi<1$. It is swept away by the wind if $\xi>1$. 
If $r_b/r_s < 1$ and $\xi < 1$,
then the explosion penetrates the wind generating zone and crosses it. While it is possible to choose parameters such that even an explosion that occurs inside the wind generating zone may not reach the very center, we will not concern ourselves with this scenario in this study.

{
\begin{table}[h]
\begin{tabular}{ll}
$M_e$ & Ejecta mass \\
$E$ & Explosion energy \\
$v$ & Wind velocity \\
$\dot{M}$ & Mass loss rate \\
$r$ & Distance hypocenter and the wind generating zone  \\
$r_w$ & Size of the wind generating zone \\
\end{tabular}
\caption[]{List of dimensional constants.}
\label{table:dimensional_constant}
\end{table}
}

\section{Numerical Simulations} \label{simulations}
We ran computer simulations using the Godunov type hydro - code RICH \citep{Yalinewich_2015}. In the Godunov scheme, the computational domain is divided into cells. Each time step, the fluxes are calculated on all the interfaces between every pair of adjacent cells. The fluxes determine how much mass, momentum and energy should be exchanged between adjacent cells. After the exchange, the primitive variables (density, pressure and velocity) are recalculated. In most simulations of this kind the mesh is static and structured. By structured we mean that the cell edges align with axes of some curvilinear coordinate system. In the RICH code, the partitioning of the computational domain into cells is achieved using the Voronoi tessellation, so the mesh does not have to be structured. Instead of starting out with cells, we assign hydrodynamic data to mesh generating points, and construct a Voronoi cell around each (a Voronoi cell around a mesh generating point is defined as the locii of all the points that are closer to that mesh generating point than to any other mesh generating point). In principle, the locations of the mesh generating points (and therefore the mesh) is allowed to change from one time step to the next. However, since the problem we are studying involves both inflow and outflow, we found it easiest to have a static (but still unstructured) mesh.

We note that in general, the simulation is carried out in dimensionless units. However, for illustrative purposes we will describe the simulation in terms of units typical of the GC.

The mesh generating points were arranged on a logarithmic spiral, such that the ratio between the cell size and its radius was constant $\frac{\Delta r}{r} \approx 0.013$. The coordinate system used was cylindrical, the computational domain was $-10 \, \rm{pc} < z < 10 \, \rm{pc}$, $0 < r < 10 \, \rm{pc}$. The GC was chosen as the origin of the axes, and a black hole was placed there. All the cells within a radius of 0.05 parsec from the black hole were designated as dummy cells that do not evolve hydrodynamically, but only absorb incident matter. This feature turned out to be unnecessary, but was kept for completeness. The outer boundary conditions were free flow, i.e. matter could freely leave the domain. 
The wind generating zone was represented by a sphere of radius 0.4 pc around the black hole (except for the series of simulations described in \ref{sec:param_sur}, where, for technical reasons, we used a radius of $5 \cdot 10^{-3} \, \rm{pc}$). Inside the wind generating zone matter was constantly added at a total rate $\dot{M} = 3 \cdot 10^{-3} \, M_{\odot}/\rm{year}$, with a small temperature (1000 K) and a radial velocity of $750 \, \rm{km/s}$. This source term was assumed to be unaffected by the hydrodynamics. We initialized the simulation with some arbitrary profile for the density, pressure and velocity. However, we let the simulation run in quiescent mode for $10^4$ years before the explosion, so as to let the initial profile be replaced by the wind profile. The supernova was introduced by the injection of matter and energy. The mass used was a typical ejecta mass $M_e = 5 \, M_{\odot}$, and the energy was a typical supernova energy $E = 10^{51} \, \rm{erg}$. We ran a series of calculations, each with a different initial separation between the explosion and the GC. In the other set of calculations we varied the ejecta mass, while keeping the initial separation constant. The supernova center was always placed on the $z$ axis, below the black hole.

\subsection{Parameter Survey} \label{sec:param_sur}
\begin{figure}[ht!]
\begin{center}
\includegraphics[width=1.1\columnwidth]{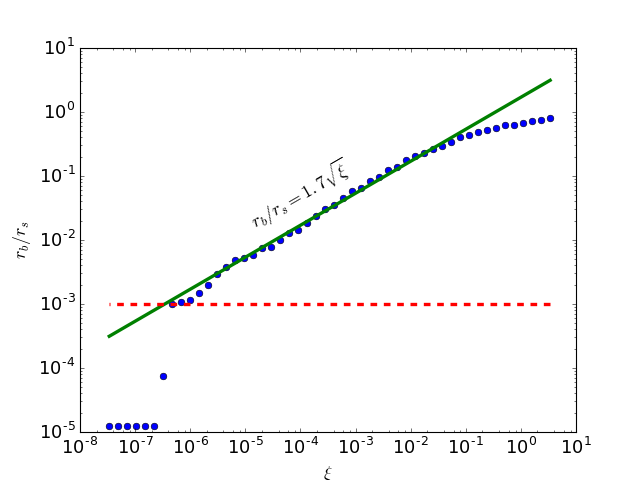}
\caption{The closest approach of the shock to the GC (normalized by the distance between GC and the wind source), as a function of $\xi$. The blue dots represent the simulation, while the solid green line represent the power law fit. The red dotted line denotes the position of the wind generating zone. 
The green line represent the fit to the analytic form $\frac{r_b}{r_s} = 1.7 \, \sqrt{\xi}$, in accordance with equation \ref{eq:rb_vs_xi}. When $r_b/r_s< 10^{-3}$ the explosion penetrates the wind generating zone. At high values of $\xi$ the ratio $r_b/r_s$ maxes out at 1 (the minimal distance cannot exceed the initial separation).
\label{fig:numeric_scaling_law}
}
\end{center}
\end{figure}

We ran a series of simulations in order to verify the scaling relation in equation \ref{eq:rb_vs_xi}. We placed the supernova progenitor at a distance of 5 pc from the GC, shrunk the wind generating zone to $5\cdot 10^{-3}$ pc, and scanned a range wind mass loss rates. In this way we changes the dimensionless parameter $\xi$, while keeping $\varpi$ and $\psi$ fixed. For each simulation we note the closest approach of the shock to the GC. The results are shown in figure \ref{fig:numeric_scaling_law}. 
The numerical results fit a relation of the form $r_b/r_s \propto \sqrt{\xi}$. However, the numerical results deviate from this relation at two extremes. At large values of $\xi$, the value of $r_b/r_s$ maxes out at 1. At very low values of $\xi$, the shock reaches the wind generating zone. If the extrapolated value of $r_b$ is well below $r_w$, then the explosion will penetrate the wind generating zone and make it all the way to the center (in our simulation it only makes it to a minimum distance of $10^{-5}$ because of the resolution).

We also performed another series of calculations keeping $\xi$ fixed at 0.2 and, instead, varied the ejecta mass, so that $\varpi$ varied between 0.01 and 1. We found that across this range, the closest approach of the shock to the GC changes by about 50\%. This result can be understood in the following way. In contrast to regular SNRs, where the transition to the Sedov phase occurs instantaneously, in this case each fluid element from the ejecta accumulates mass at a different rate, and therefore, reaches its Sedov phase at a different time. In the relevant range of $\varpi$ values, the fluid elements moving toward the wind source transition to the Sedov phase prior to their arrival to minimal distance, so their original mass is less important. We note that in astrophysical scenarios the variations in ejecta mass are smaller than the range we considered.

The third dimensionless parameter is $\psi$. We can see from figure \ref{fig:numeric_scaling_law} that there are two regimes. In the first regime, the closest approach of the shock, $r_b$, is larger than that of the wind generating zone, $r_w$, and $r_b$ is independent of $r_w$. In the other regime, the shock penetrates the wind generating zone and makes it all the way to the center. The size of the wind generating zone determines, obviously, the transition between these two regimes.

\begin{figure}[ht!]
\begin{center}
\includegraphics[width=1.1\columnwidth]{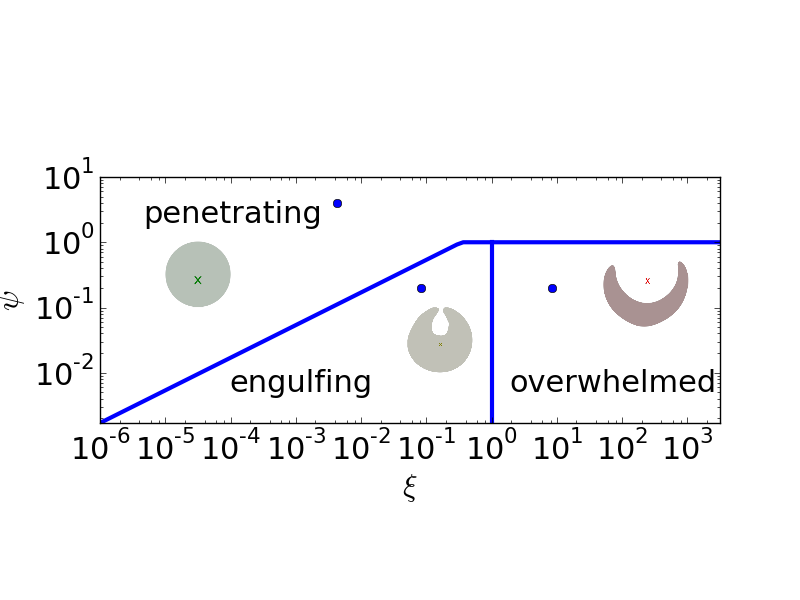}
\caption{
A ($\xi,\psi$) phase diagram of  the different explosion morphologies (for a fixed $\varpi$). The definitions of these dimensionless variables is given in section \ref{sec:analytic_estimates}. The blue lines are the boundaries between the different domains. The blue dots represent the values used in the numerical simulations. A schematic illustration of the explosion morphology
is depicted in each region. 
\label{fig:phase_diagram}
}
\end{center}
\end{figure}

The discussion in this section shows that, for the astrophysical scenarios that we consider here, the morphology is determined by two dimensionless variables: $\psi$ and $\xi$. Figure \ref{fig:phase_diagram} shows a schematic ``phase diagram" for the different morphology, i.e. the regions in parameter space where each morphology manifests itself.

\subsection{A Penetrating Explosion} \label{sec:penetrating_explosion} 
We placed the progenitor in the wind generating zone, at a distance of $0.1 \, \rm{pc}$ from the GC. The corresponding dimensionless variables are $\xi = 4.25\cdot 10^{-3}$, $\psi = 4$ and $\varpi = 2.45$.

\begin{figure}[ht!]
\begin{center}
\includegraphics[width=1.1\columnwidth]{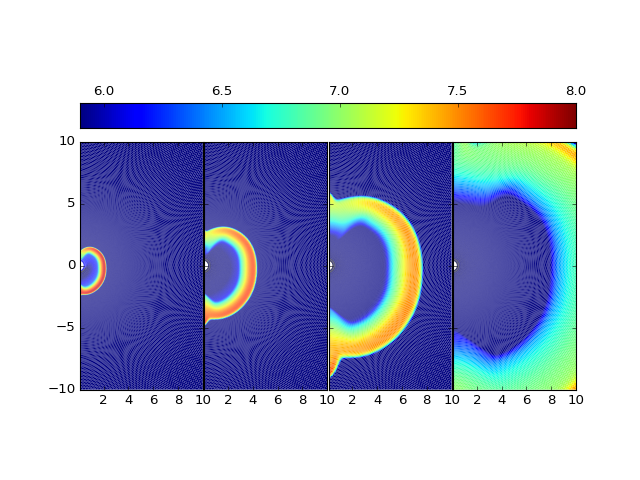}
\caption{Temperature snapshots from a simulation where the explosion penetrates the wind generating zone. The color scale is $\log_{10}\left( T/1 \, \rm{K}\right)$. The ages of the supernova in each image are (from left to right) 660, 1500, 3000 and 6000. The r and z axes are in parsecs.
\label{fig:magnetar_sim}
}
\end{center}
\end{figure}

A selection of four snapshots of the temperatures is shown in figure \ref{fig:magnetar_sim}. As expected, the ejecta expels all matter from the vicinity of the GC within a few hundred years. However, after a few thousand years, while the ejecta is still propagating outwards, the shocked matter inside is being pushed outside by the new cold stellar wind.

\subsection{An Engulfing Explosion} \label{sec:engulfing_explosion}
\begin{figure}[ht!]
\begin{center}
\includegraphics[width=1.1\columnwidth]{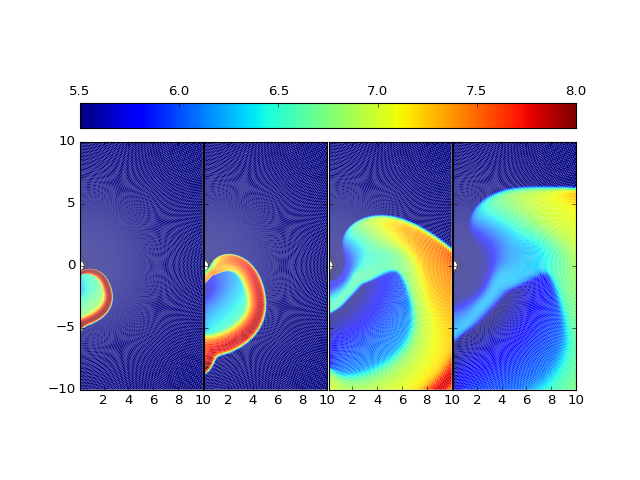}
\caption{Temperature snapshots from the simulation where the explosion engulfs, but does not penetrate the wind generating zone. The color scale is $\log_{10}\left( T/1 \, \rm{K}\right)$. The ages of the supernova (left to right) are 660, 1500, 4000 and 6000 years.
\label{fig:sgr_east}
}
\end{center}
\end{figure}

If we increase the initial distance between the explosion and the GC, at some point the ejecta won't penetrate the wind generation zone, but will rather engulf if.  To achieve this, we choose the initial distance between the explosion and the center of the wind source to be 2 parsec. The corresponding dimensionless variables are 
$\xi = 8.51\cdot 10^{-2}$, $\psi = 0.2$ and $\varpi = 2.45$ (same as in the previous section). The explosion shock wave is at its closest approach to the wind generating zone at $t = 650 \, \rm{year}$. Incidentally, the closest approach turns out to be just the radius of the wind generating zone $r_w = 0.4 \, \rm{pc}$. 
A selection of snapshots can be seen in figure \ref{fig:sgr_east}.

The simulation shows that the shock wave does not enclose the wind generating zone from all sides. Within $10^4$ years the shocked material has already been pushed away, and the environment in the inner parsecs returns to its state prior to the explosion.

\begin{figure}[ht!]
\begin{center}
\includegraphics[width=1.1\columnwidth]{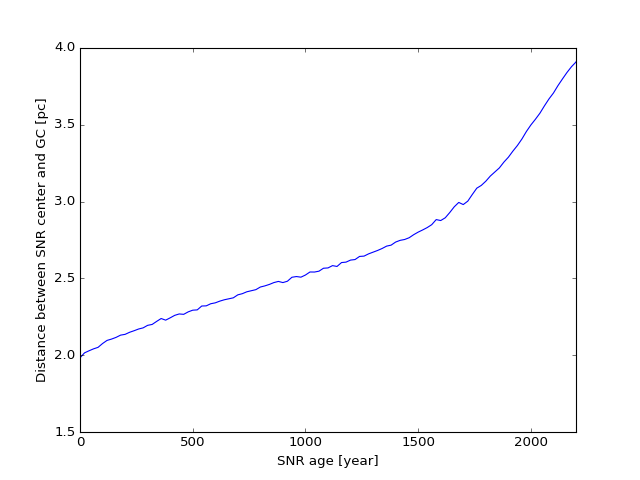}
\caption{Distance of the center of an engulfing SNR, as a function SNR age, calculated from the simulation. Until 1500 years the drift is at the velocity of the wind, and from 1500 onward the drift is at half the velocity of the SNR shock.
\label{fig:drift_history}
}
\end{center}
\end{figure}

One phenomenon that can be seen in the images is that the apparent center of the explosion drifts away from the center. While the radius of the SNR is much smaller than the distance to the GC, the whole SNR drifts at the velocity of the wind. At later times, a part of the shocked gas approaches the GC and stalls, while the other end continues to move away from the GC. At this stage the drift velocity of the center is half of the velocity of the shock moving away from the GC, which still faster than that of the wind. This behavior, where the center drifts slowly at first, and later accelerates, is demonstrated in figure \ref{fig:drift_history}.

\subsection{An Overwhelmed Explosion} \label{sec:overwhelmed_explosion}
\begin{figure}[ht!]
\begin{center}
\includegraphics[width=1.1\columnwidth]{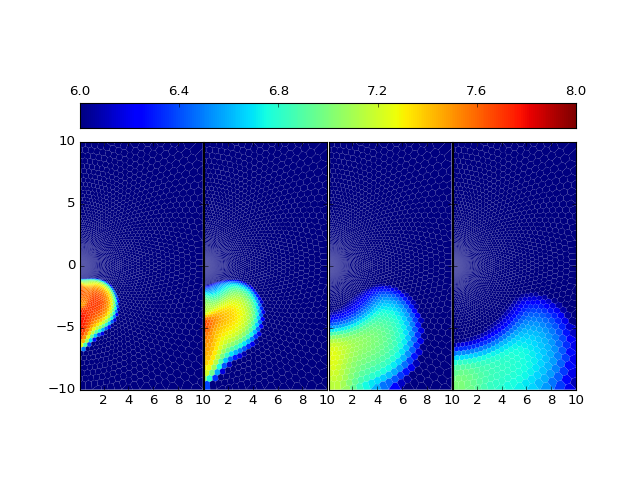}
\caption{Temperature snapshots of the overwhelmed explosion simulation. The color scale is $\log_{10}\left( T/1 \, \rm{K}\right)$. The ages of the supernova (left to right) are 1000, 2000, 4000 and 6000 years. The r and z axes are in parsecs.}
\label{fig:overwhelmed}
\end{center}
\end{figure}

In this simulation we increased the mass loss rate of the wind by a factor of 100 (to $\dot{M} = 3\cdot 10^{-1} \, M_{\odot}/\rm{year}$), and set the initial separation between the explosion and the GC to 2 pc, so as to exceed the threshold in equation \ref{eq:st_radius}. We recall that according to the analytic analysis, such explosion would never engulf the wind generating zone, but would be swept away. The corresponding dimensionless variables are $\xi = 8.51$, $\psi = 0.2$ and $\varpi = 2.45$ (same as in the previous section).
Figure \ref{fig:overwhelmed} shows a selection of temperature snapshots. This simulation supports our analytic estimates.

Like the engulfing explosion, the apparent center of this explosion also drifts outwards, away from the center. In this case, however, the drift is only at the speed of the wind, rather than that of the ejecta.

\vspace{5mm}

\section{Justification of our Model} \label{sec:justification}
So far we've presented a simplified model. The model focuses only on the hydrodynamic interaction of the ejecta with the wind and ignores other effects.  Before turning to observational implications of this model we discuss several phenomena that were omitted and assess their importance.

\subsection{Gravity}
The super massive black hole at the center attracts the surrounding matter with its gravitational field. The gravitational effects are important when the velocity of the flow is equal or lesser than the Keplerian velocity. The gravitational effects become unimportant beyond the Bondi radius
\begin{equation}
r_{bo} = \frac{G M_{bh}}{v^2} \approx 0.1 M_{bh,6} v_3^{-2} \, \rm{pc}
\end{equation}
$M_{bh}$ is the mass of the black hole and $M_{bh,6} = M_{bh}/10^6 M_{\odot}$.
This radius is much smaller than the other length scales in the problem, so gravitational effects can be neglected. The orbital motion of the exploding star or the surrounding matter can also be neglected beyond this radius.

In other galaxies, SMBH masses can be as high as $10^{10} M_{\odot}$. Since the Bondi radius increases linearly with the mass, then it can be as large as 30 pc. In such systems the effects of gravity can no longer be neglected.

\subsection{Radiative Losses} \label{sec:radiative_losses}
Radiative cooling becomes important when the cooling timescale is shorter than the age of the explosion. We denote this time by $t_r$. By $t_r$ most of the original energy of the explosion would be gone. We make the simplifying assumption that by $t_r$ the radius of the explosion is much larger than the initial distance between the explosion and the GC. Therefore, the explosion can be considered concentric (see section \ref{sec:concentric_explosion}). The onset of radiative cooling cannot occur during the ejecta dominated phase, because at that point most of the energy is kinetic. 
The cooling time for the Sedov - Taylor phase is
\begin{equation}
t_r = \frac{E^4 v^7 m_p^6}{\Lambda^3 \dot{M}^7}
\end{equation} 
where $\Lambda$ is the cooling function. The condition for radiative cooling to begin during the Sedov - Taylor phase is
\begin{equation}
1 > \left(\frac{t_r}{t_{dr}} \right)^{1/3} \approx 32 \, E_{51} v_3^3 \Lambda_{-21}^{-1} \dot{M}^{-2}_{-3}
\label{eq:radiative_timescale}
\end{equation}
where $\Lambda_{-21} = \Lambda / 10^{-21} \, \rm erg \cdot cm^3 / s$.
In the last, coasting, phase of the concentric explosion the cooling time scale increases linearly with the age. Hence, if the onset of radiative cooling did not occur before, it will not occur during this phase. Overall, we get that radiative cooling can begin only during the Sedov - Taylor phase.

For our canonical values, in the GC, inequality \ref{eq:radiative_timescale} does not hold, so radiative energy depletion does not occur. 
However, inequality \ref{eq:radiative_timescale} can be satisfied in the centers of other galaxies. In particular it is possible that the wind velocity is smaller by a factor of 3, or the mass loss rate is larger by a factor of 6. In fact, recent studies of the radio emission from  TDE ASSASN 14li \citep{2016arXiv160202824K} showed that in the host GC, at the same distance from the central black hole, the density is about 10 times higher than in our GC. This could either mean a higher mass loss rate or a slower velocity. In each case inequality \ref{eq:radiative_timescale} will be satisfied.

If radiative cooling starts in the Sedov Taylor phase and the explosion is spherical, then after $t_r$ the snowplow phase begins. In this phase all the thermal energy is immediately radiated away. The evolution of the explosion is determined by the conservation of radial momentum rather than global energy conservation 
\begin{equation}
R \approx \left( \frac{E v t_r^2}{\dot{M}} \right)^{1/3} \left( \frac{t}{t_r}\right)^{1/2} \,.
\end{equation}

Things are much more complicated if at $t_r$ the radius is not much larger than the initial distance between the explosion and the GC. Not only is the explosion non spherical, but different parts can enter the radiative phase at different time. However, in order for radiative cool to start at the early stages of the explosion requires a radical values of the parameters, so we will not discuss them in this work.

\subsection{Molecular Clouds}
Scattered throughout the GC are molecular clouds with densities in excess of $10^4 \, \rm{cm}^{-3}$. Their sizes vary, but the biggest can reach a few parsec. Their influence on the flow has been studied using a SPH simulation \citep{rockefeller_et_al_2005}, and it was found that the flow overtakes these dense clumps without imparting momentum and energy to them. We can therefore safely ignore them in this work.

\subsection{The Wind swept Environment}
In the analysis so far we assumed that the $r^{-2}$ wind profile extends indefinitely. Obviously, this assumption has to break at some point, because the average galactic ISM density is much larger than what the extrapolation of the $r^{-2}$ profile would yield. Current observations are not sensitive enough to pinpoint the radius where the transition happens, but one can infer from them bounds on the transition radius.

The measured density of ionized gas at distances between tens and hundreds of parsecs from the GC (i.e. the central molecular zone, hereafter CMZ) is of the order of $10 \, \rm{cm}^{-3}$ \citep{2003astro.ph..1598C}. It was also found that the density distribution is not spherically symmetric. The density declines on a length scale of about 150 parsec along the galactic disc, and on a length scale of 30 parsec perpendicular to it \citep{langer_pineda_2015}. Furthermore, the motion of the gas is not radial, but turbulent, with typical velocities much smaller than that of the GC wind \citep{langer_et_al_2015}. It is currently unclear how, and at what radius the density profile in the wind swept environment of the GC connects to the turbulent environment of the central molecular zone. The 
lower bound on this distance is 5 parsec, 
because at that distance the density of the wind is equal to the density of the CMZ. Therefore, it is safe to assume that the density profile due to the wind persists to a radius of at least 5 parsec.

\section{SNRs near our Galactic Center} \label{sec:snr_gc}
\subsection{Observations}
The close proximity (inner few parsecs) of the GC is a complex environment. At the very center lies a super massive black hole - SGR A*. Its mass is estimated at $4.3 \cdot 10^6 M_{\odot}$ \citep{gillessen_et_al_2009}, and the inferred accretion rate is about $10^{-6} \, M_{\odot}/\rm{year}$, much smaller than the Eddington limit ($10^{-2} \, M_{\odot} / \rm{year}$). The environment around it contains plasma, dust, cold molecular clouds and stars. The plasma densities and temperatures were measured by Chandra \citep{baganoff_et_al_2003}, and it was found that the density ranges between about $130 \, \rm{cm}^{-3}$ at a distance of 0.04 parsec, to $30 \, \rm{cm}^{-3}$ at 0.4 parsec, and the temperature is of the order of a few keV. The amount, composition and grain size distribution is currently not fully known. Even though its mass may be negligible compared to the other components, dust is the chief source of extinction in the GC in the infrared, visible and ultraviolet ranges. There are also numerous scattered clouds, most with densities in excess of $10^{4} \, \rm{cm}^{-3}$ and with a low filling factor. The largest collection of dense clumps is in the circum - nuclear disk. This ring - like structure whose thickness is about 1.5 parsec, closest point to SGR A* is at a distance of about 2 parsec, and the farthest is 10 parsec. Finally, there are about $10^6$ stars within a few parsecs of SGR A*. At about 3 parsec the combined masses of the stars equals that of SGR A*. Some of those stars emit winds. A small portion of that wind is accreted onto SGR A*, but most of it escapes the GC complex. The typical velocity of the wind is around 700 km/s, and the net mass loss rate from the GC due to winds is about $3\cdot 10^{-3} M_{\odot}/ \rm year$ \citep{rockefeller_et_al_2004}. This wind determines the density and temperature profiles of the plasma \citep{quataert_2004}. The same wind may also be the source of the dust.

The GC is enshrouded in a luminous blob, thought to be a supernova remnant, called SGR East. The center of this blob lies at a distance of 2 parsec from SGR A*. SGR East is not quite spherical (the ratio between the major and minor axes is about 2:3), and the farthest point in this blob is about 5 parsec away from the center of the blob. 

More recently, another SNR candidate has been identified in the SGR A's bipolar radio/X-ray Lobes \citep{ponti_et_al_2015}. This SNR appears symmetric around the GC, its center almost coincides with SGR A* and its radius is approximately 12 pc. 

\begin{figure}[ht!]
\begin{center}
\includegraphics[width=1.1\columnwidth]{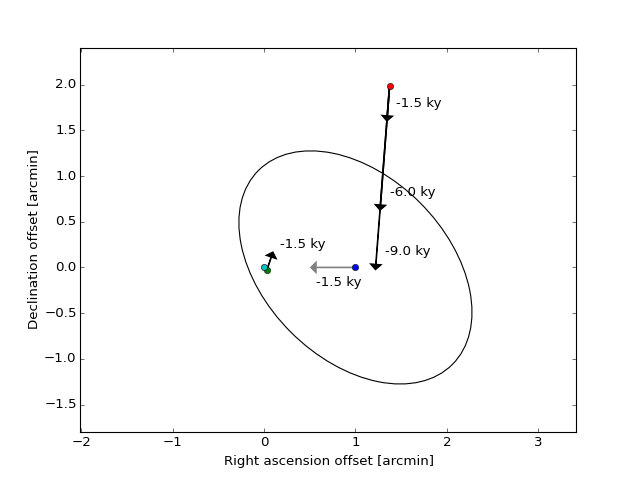}
\caption{Schematic illustration of current and extrapolated past positions of GC objects.  The cyan dot is SGR A*, the green dot is the magnetar SGR J1745-2900, the blue dot is the center of SGR east (black ellipse) and the red dot is the cannonball. The arrows show where the object would have been in the past, if they were moving at constant velocity. This extrapolation is valid for the cannonball, but not for the magnetar, which is most likely bound to SGR A*. In the case of the center of SGR East the arrow marks the maximal apparent drift, while the actual apparent drift can be smaller because of inclination.}
\label{fig:gc_complex}
\end{center}
\end{figure}

Two pulsars have been detected close to SGR A*: CXOGC J174545.5-285829 (a.k.a the cannonball) \citep{nynka_et_al_2013} and SGR J1745-2900 \citep{kaspi_et_al_2014} (which is also a Magnetar). The typical age of both is $9 \cdot 10^{3}$ years \citep{mori_et_al_2013, kaspi_et_al_2014}. The locations of the SNRs and pulsars are presented schematically in figure \ref{fig:gc_complex}. The cannonball is moving away from the GC at a velocity of about $500 \, \rm{km/s}$ \citep{nynka_et_al_2013}. The exact trajectory of the magnetar SGR J1745-2900 is not yet known at a very high accuracy, but its current proper velocity is about $230 \, \rm{km/s}$, and is most likely bound to SGR A* \citep{bower_et_al_2016}. 

The GC complex is surrounded by a dense gas disc \citep{launhardt_et_al_2002}. The disc extends, along the galactic plane, to a distance of about 200 pc. The height of the disc, i.e. its extent perpendicular is 50 pc. The average electron density at the center of the disc is $10 \, \rm{cm}^{-3}$ \citep{jm_cordes_et_al_1993}.

\subsection{Length scales in the Galactic Center}

The GC of the Milky Way has a $4.3 \cdot 10^6 M_{\odot}$ black hole, a wind generating zone of radius $r_w \approx 0.4 \, \rm{pc}$, wind velocity $v \approx 700 \, \rm{km/s}$ and net mass loss rate $\dot{M} \approx 3 \cdot 10^{-3} M_{\odot} / \, \rm{year}$. The Bondi radius for the wind is about 0.03 pc. This is much smaller than all other length scales and hence the gravitational field of the black hole can be neglected. 
A typical supernova explosion with $E \approx 10^{51} \, \rm{erg}$ and mass $M_e \approx 5 M_{\odot}$, 
explosions occurring at distances below 2 pc will penetrate the wind generating zone. The morphology of such explosions is shown as a green circle in figure \ref{fig:schematic}, and its simulation is discussed in section \ref{sec:penetrating_explosion}. Explosions at distances between 2 and 50 pc will engulf the wind generating zone, but won't penetrate it. This morphology is shown as a yellow horseshoe in figure \ref{fig:schematic}, and its simulation is discussed in section \ref{sec:engulfing_explosion}. At distances larger than 50 pc the explosion will be overwhelmed. This morphology is shown as a red crescent in figure \ref{fig:schematic}, and the its simulation is discussed in section \ref{sec:overwhelmed_explosion}.

\subsection{The Missing Relic of SGR East}
In the ejecta dominated phase the radius grows linearly with time $r \approx t \sqrt{E/M}$. Using our canonical values $E = 10^{51} \, \rm{erg}$, $M = 5 M_{\odot}$ yields that a typical SNR in the GC attains a radius of $5 \, \rm{pc}$ within $1500 \, \rm{years}$. This result is demonstrated by the numerical simulations shown in figure \ref{fig:sgr_east}. This result is in agreement with an earlier simulation by \citet{rockefeller_et_al_2005}. This SPH simulation included details, like molecular clouds, that cannot be captured in our simulations. The agreement between the simulations shows that the results do not depends on the numerics or on the details of the density structure around the GC. 

The simulations also show that the apparent center of an engulfing explosion drifts away from the GC at a rate of 1 pc every 1400 years. The apparent drift can be smaller because of inclination. 

In an earlier study  \citep{mezger1989continuum} the age of SGR East was estimated at $10^4$ years. They suggested that the high metallicity in SGR East was because the explosion occurred inside a dense molecular cloud and assumed that throughout its evolution, the SNR expanded into an ambient medium with a density of about $100 \, \rm{cm}^{-3}$. However, recent Chandra measurements \citep{baganoff_et_al_2003} showed that such high densities occur only within a radius of $0.04 \, \rm{pc}$, and that at a radius of $0.4 \, \rm{pc}$ the density drop by an order of magnitude.

Following this ($10^4$ year) age estimate, it was suggested that SGR East and CXOGC J174545.5-285829 (the cannonball) are associated \citep{sangwook_et_al_2005,zhao_et_al_2013}. The cannonball is a runaway neutron star, currently at a distance of about 5 pc from the GC, and it is moving away from it at a velocity of $500 \pm 100 \, \rm{km/s}$. Retracing its trajectory ten thousand years into the past puts it very close to the current apparent center of SGR East.

The new age estimate poses a problem for the association of SGR East with the cannonball. According to this estimate, the neutron star would not have had time to travel to its current location from where the progenitor used to be. This difficulty is further aggravated by the drift of the apparent center of SGR East, because it means the progenitor was even closer to the GC (see figure \ref{fig:gc_complex}).

One could argue that by stretching the parameters it may be possible to extend the age estimate according to our model. However, this requires the velocity of the ejecta to decrease from about $3000 \, \rm{km/s}$ (velocity of the ejecta for our canonical values) to $500 \, \rm{km/s}$. There are no known supernovae with such a low ejecta velocity.

It is impossible to construct a Keplerian orbit that has its focus at SGR A*, passes through the current location of SGR J1745-2900 and the center of SGR East 1500 years ago, and that the object moving along it can cover the distance between the two point within 1500 years, moving at velocities below the magnetar's current measured velocity. Therefore SGR J1745-2900 cannot be associated with SGR East as well. 

Our age estimate for SGR East may also explain the abundance of dust close to the GC \citep{lau_et_al_2014}. It is currently thought that dust forms during the ejecta dominated phase of a supernova, but some of it is destroyed in the transition to the Sedov Taylor phase. According to our age estimate, SGR East is still in the ejecta dominated phase, so this may explain the abundance of dust inside.

\subsection{The Association between SGR A's bipolar Radio/X-ray Lobes and SGR J1745-2900}
According to our simulations, a SNR can attain its radius within about 6000 years. This is relatively close to the estimated ages of both pulsars. Since the the SNR is symmetric around the galactic plane, and SGR A* is close to its center, then its corresponding relic could have easily been SGR J1745-2900. It can't be the cannonball, because an explosion that occurred where the cannon was 6000 years ago (see figure \ref{fig:gc_complex}) would result in an engulfing explosion (see figure \ref{fig:schematic}, section \ref{sec:engulfing_explosion}) that is asymmetric at this stage (as can be seen in figure \ref{fig:sgr_east}), whereas the new SNR candidate is symmetric around the galactic plane \cite{ponti_et_al_2015}.

In contrast to SGR East, this SNR exploded very close to SGR A*, so it could have disrupted the accretion to it. The accretion pause would only last a few hundred years, until new matter from the stellar winds reached SGR A* and fed the SMBH again.

\subsubsection{The Missing SNR of the Cannonball}
In the previous subsections we showed that the cannonball cannot be associated with either observed SNR. This means that its associated supernova has not been detected. This SNR can be 9000 years old, so it should be bigger than SGR A's bipolar radio/X-ray Lobes. We discuss two possible scenarios that may explain why the cannonball SNR has not been observed. In the first scenario, the wind profiles extends further than the radius of the SNR, so the density at the shock front keeps declining. In this case, the SNR  will dim, to the point that it's impossible to distinguish from the background. In the second scenario, at some point the ISM density stops decreasing with distance, and remains at the average CMZ density $10 \, \rm{cm}^{-3}$. At such high densities the cooling time is
\begin{equation}
t_c \approx 8 \cdot 10^3 E_{51} \Lambda_{-21}^{-1} n_{1}^{-2} R_{1}^{-3} \, \rm year
\end{equation}
where $n_1 = n/ 10 \, \rm cm^{-3}$ is the ambient density and $R_1 = R/10 \, \rm pc$ is the radius of the SNR.
Therefore, in this scenario the energy will be dissipated within a few thousand years. In both scenarios the SNR will not be observable after 9000 years. More detailed observational information on the distribution of ISM density in the CMZ will allow us to decide between these two scenarios.

\section{Discussion} \label{sec:discussion}
We studied the evolution of supernova remnants near the GC. We saw that such SNRs evolve differently from regular SNRs because of the strong wind emanating from the stellar cusp close to the GC. The explosion can behave in one of three ways. An explosion very close to the GC will be spherical. An explosion very far from the GC will be overwhelmed and swept away. An explosion in between will engulf the GC, but will not reach it. We reproduced these behaviors using a two dimensional numerical simulation. 

We have shown that the age of SGR East is around 1500 years, in accordance with \citet{rockefeller_et_al_2005}. We have also shown that the apparent center of SGR east drifts away from the GC at a rate of about 1 pc every 1400 years. The projected distance between the center of SGR East and the cannonball today is about 5 pc, and the cannonball is moving away from the center of SGR East at 500 km/s. 1500 years ago, the distance between them would have been 4 pc. The separation is even larger when the drift is taken into account.

The other neutron star (SGR J1745-2900) is located on the same side of SGR A* as the center of SGR East at a projected distance of 0.1 pc from SGR A*. It moves at a velocity of 200 km/s in a direction perpendicular to the line connecting it to the center of SGR East (see figure \ref{fig:gc_complex}). There is no Keplerian of orbit that connects the current position of SGR J1745-2900 and the past position of the center of SGR East, and that moving along it SGR J1745-2900 can travel between the two points in 1500 years. Hence, SGR East cannot be associated with SGR J1745-2900 as well. The lack of associated compact object can be explained in two ways. One option is that the object exists, but has not been observed. The second is that a compact object was never created. This could happen in the case of type Ia and pair instability supernovae.

We also show that the cannonball cannot be associated with SGR A's bipolar radio/X-ray Lobes. The age of this SNR is estimated at 6000 years, and the distance between SGR A* and the cannonball 6000 years ago was 4 pc. At that distance the explosion would not have been symmetric with respect to reflection around SGR A*, while this SNR is symmetric. The only valid association is between this SNR and SGR J1745-2900.

In addition, this study has two other interesting implications. First, Supernovae that occur at distances larger than 2 parsec from the GC will not penetrate the wind generating zone, and therefore, would not disrupt the accretion unto SGR A*. Second, the new age estimate may explain the abundance of dust around the GC \citep{lau_et_al_2014}. It is currently thought that dust forms during the ejecta dominated phase of a SNR, but some of it may be destroyed in the transition to the Sedov Taylor phase. According to the the age estimate in this work (and also in \citet{rockefeller_et_al_2005}), SGR East is still in the ejecta dominated stage, and this explains why there is so much dust inside the shocked region. 

The research was supported by a grant from the Israel Space Agency and by the ISF-CHE I-Core center of excellence for research in Astrophysics and an ISF grant.

\bibliographystyle{yahapj}
\bibliography{almogyalinewich.bib}

\end{document}